\documentclass[twocolumn,secnumarabic,amssymb, nobibnotes, aps, prd]{revtex4}

\usepackage{graphicx}
\newcommand{\be}{\begin{equation}} 
\newcommand{\ee}{\end{equation}} 
\newcommand{\bea}{\begin{eqnarray}} 
\newcommand{\eea}{\end{eqnarray}} 
\newcommand{\bm}{\begin{mathletters}} 
\newcommand{\eml}{\end{mathletters}} 
\newcommand{\nn}{\nonumber\\} 
\newcommand{\oh}{\frac{1}{2}}

\begin{document} 

\title{Universal decay rule for reduced widths}

\author{D.S. Delion$^{1,2}$}
\affiliation{
$^1$ Horia Hulubei National Institute of Physics and Nuclear 
Engineering, \\ 
407 Atomi\c stilor, Bucharest-M\u agurele, 077125, Rom\^ania \\
$^2$ Academy of Romanian Scientists,
54 Splaiul Independen\c tei, Bucharest, 050094, Rom\^ania}

\begin{abstract}
Emission processes including $\alpha$-decay, heavy cluster decays, 
proton and di-proton emission are analyzed in terms of the well known  
factorisation between the penetrability and reduced width. 
By using a shifted harmonic oscilator plus Coulomb cluster-daughter 
interaction it is possible to derive a linear relation between the 
logarithm of the reduced width squared and the fragmentation potential, 
defined as the difference between the Coulomb barrier and Q-value.
This relation is fulfilled with a good accuracy for transitions 
between ground states, as well as for most $\alpha$-decays
to low lying $2^+$ excited states.
The well known Viola-Seaborg rule, connecting half lives 
with the Coulomb parameter and the product between fragment charge 
numbers, as well as the Blendowke scalling rule connecting the
spectroscopic factor with the mass number of the emitted cluster,
can be easily understood in terms of the fragmentation potential.
It is shown that the recently evidenced two regions in the dependence of 
reduced proton half-lives versus the Coulomb parameter are directly 
connected with the corresponding regions of the fragmentation 
potential.

\end{abstract}


\maketitle

\section{Introduction} 
\label{sec:intro} 
\setcounter{equation}{0} 
\renewcommand{\theequation}{1.\arabic{equation}} 

The family of emission processes triggered by the strong interaction contains
various decays, namely particle (proton or neutron) emission,
two-proton emission, $\alpha$-decay, heavy cluster emission and
binary or ternary fission.
There are also other nuclear decay processes induced by electromagnetic
($\gamma$-decay) or weak forces ($\beta$-decays).
The purpose of this work is to investigate only the first type of 
fragmentation, where the emitted fragments are left in ground or 
low-lying excited states.
They are called cold emission processes and are presently among important
tools to study nuclei far from the stability line.
Nuclei close to the proton drip line are investigated through proton emission,
while the neutron drip line region is probed by cold fission processes.
On the other hand superheavy nuclei are exclusivelly detected by 
$\alpha$-decay chains \cite{Gam05}.
Actually the first paper in theoretical nuclear physics 
applying quantum mechanics \cite{Gam28} was
devoted to the description of the $\alpha$-decay in terms of the penetration
of a preformed particle through the Coulomb barrier.

There are two goals of this paper. The first one is to explain the well 
known Viola-Seaborg rule  \cite{Vio66}, valid for all kinds of cold 
emission processes.
It turns out that it is possible to give a simple interpretation
of this rule in terms of two physical quantities, namely the Coulomb 
parameter, connected with the penetrability, and the fragmentation 
potential, connected with the reduced width. An universal linear
dependence between the logarithm of the reduced width and
fragmentation potential will be derived.
It will be shown that this interpretation is valid not only for 
transitions  between ground states, but also for transitions to excited 
states.
On the other hand, the scalling dependence of spectroscopic factors in
heavy cluster decays versus the mass numbers of the emitted cluster
can be also understood in terms of the fragmentation potential.

\section{Experimental decay rules}
\label{sec:redwid}
\setcounter{equation}{0} 
\renewcommand{\theequation}{2.\arabic{equation}} 

Let us consider a binary emission process $ P\rightarrow D+C$
from a parent ($P$) to the daughter nucleus ($D$) and 
the lighter cluster ($C$), which can be in particular an 
$\alpha$-particle or a proton.
The total decay width is the sum of partial decay widths
corresponding to different angular momenta, given by \cite{Del06}
\bea
\label{gam} 
\Gamma_{l}=
2P_l(E_l,r)\gamma^2_l(\beta,r)~,
\eea 
where it was introduced the standard penetrability and 
reduced width squared \cite{Lan58}
\bea 
\label{Pg}
P_l(E_l,r)&=&\frac{\kappa_l r}
{\left|H^{(+)}_l(\chi_l,\kappa_lr)\right|^2}~,
\nn 
\gamma^2_l(\beta,r)&=& 
\frac{\hbar^2}{2\mu r}|s_l^{(int)}(\beta,r)|^2 ~. 
\eea 
The outgoing spherical Coulomb-Hankel function 
$H^{(+)}_l(\chi_l,\kappa_lr)$ depends upon two 
variables, namely the Coulomb (or twice the Sommerfeld) parameter
\bea
\label{Somer}
\chi_l=\frac{Z_DZ_Ce^2}{\hbar v_l}~,
\eea
and the reduced channel radius $\rho_l=\kappa_l r$. Here 
$v_l=\hbar\kappa_l/\mu$ and $\kappa_l=\sqrt{2\mu E_l}/\hbar$ 
are the asymptotic relative velocity and momentum between the emitted
fragments, respectivelly, in terms of the reduced mass 
of the daughter-cluster system $\mu$. It is also defined
the center of mass (cm) channel energy $E_l=Q-E_l^{(ex)}$ 
of emitted  fragments, in terms of the difference
between the total energy (Q-value) and the excitation energy of 
the daughter nucleus $E_l^{(ex)}$.
The internal component $s_l^{(int)}(\beta,r)$ at a certain radius $r$ 
inside the Coulomb barrier, is for deformed  emitters a superposition 
of different Nilsson components multiplied by the propagator matrix 
\cite{Del06}, depending on deformation parameters $\beta$, i.e.
\bea
\label{Kll}
s_l^{(int)}(\beta,r)=\sum_{l'}K_{ll'}(\beta,r)f^{(int)}_{l'}(r)
\eea
For spherical emitters with $K_{ll'}=\delta_{ll'}$ it coincides with the 
wave function component $f^{(int)}_l(r)$.

The half life is defined by the inverse of the total decay width, i.e.
\bea
T=\frac{\hbar ln~2}{\Gamma}~.
\eea

\begin{figure}[ht] 
\begin{center} 
\includegraphics[width=9cm]{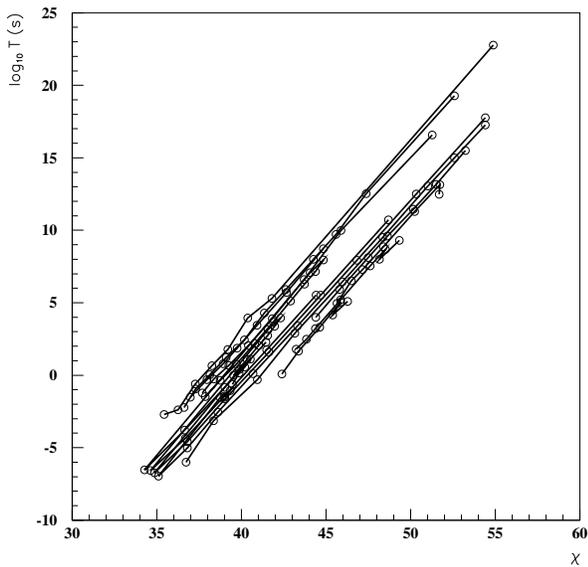} 
\vspace{-1cm} 
\caption{\it Logarithm of half lives for $\alpha$-decays from even-even nuclei
versus Coulomb parameter (\ref{Somer}).
Differents lines connect decays from nuclei with the same charge number.} 
\label{fig1} 
\end{center} 
\end{figure} 

Inside the Coulomb barrier the complex Coulomb-Hankel function 
practically  coincides with the real irregular Coulomb function and has 
a very simple WKB ansatz \cite{Del06}
\bea 
\label{Hplus}
H^{(+)}_l(\chi,\rho)
&\approx&\frac
{exp\left[\chi\left(arccos\sqrt{x}-\sqrt{x(1-x)}\right)\right]}
{\left(\frac{1}{x}-1\right)^{1/4}}C_l
\nn
&\equiv& H^{(+)}_0(\chi,\rho) C_l
\eea
where, with the external turning point $r_e=Z_1Z_2e^2/E$ and 
barrier energy $V_0=Z_1Z_2e^2/r$, there were introduced the following 
notations
\bea
x&=&\frac{\rho}{\chi}=\frac{r}{r_e}=\frac{E}{V_0}
\nn
C_l&=&exp\left[\frac{l(l+1)}{\chi} 
\sqrt{\frac{\chi}{\rho}-1}\right]~.
\eea 
There were considered here for simplicity transitions between
states with the same angular momentum $l$, as for instance
$\alpha$-decays or proton emission processes between ground states.
Thus, the logarithm of the so-called reduced half-life, corrected by the 
exponential centrifugal factor squared $C_l^2$, defined by the second 
line of this relation, i.e.
\bea
\label{Tred}
log_{10}T_{red}&=&log_{10}\frac{T}{C_l^2}
\nn&=&
\log_{10}\frac{\hbar ln~2}{C_l^2}
-log_{10}~2P_l-log_{10}\gamma_l^2,
\eea
should be proportional with the Coulomb parameter, i.e.
\bea
\label{Tchi}
log_{10}T_{red}=a_0\chi+b_0~,
\eea
Notice that in most of decay processes between ground states one has
boson fragments, with zero angular momentum, i.e. $C_l=1$.
The case with $l\neq 0$ is connected with fermions,
i.e. proton emission, where $C_l\neq 1$.

\begin{figure}[ht] 
\begin{center} 
\includegraphics[width=9cm]{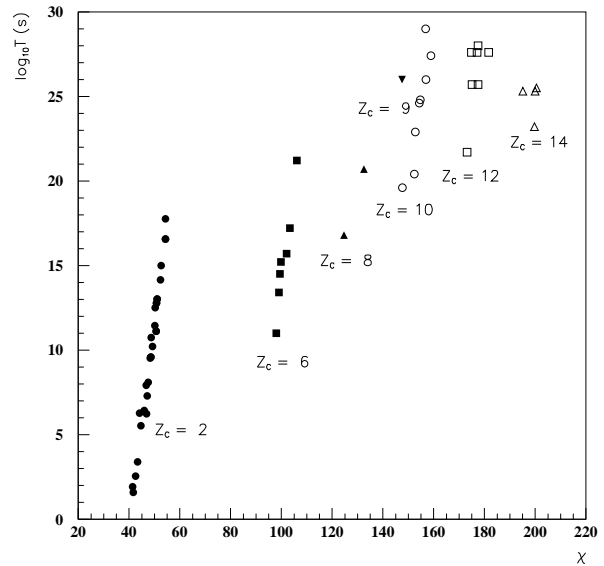} 
\vspace{-1cm} 
\caption{\it Logarithm of half lives for heavy cluster decays and the
corresponding $\alpha$-decays from the same mother nuclei
versus Coulomb parameter (\ref{Somer}).
Differents symbols denote charge number of the emitted cluster.} 
\label{fig2} 
\end{center} 
\end{figure} 

The above relation is also called Geiger-Nuttall law, discovered in 1911
for $\alpha$-decays between ground states (where the angular momentum
carried by the $\alpha$-particle is $l=0$).
The explanation of this law was given by G. Gamow in 1928 \cite{Gam28}, 
in terms of the quantum-mechanical penetration of the Coulomb barrier, 
i.e. the first line of Eq. (\ref{Pg}).
It is characterized by the Coulomb parameter, which is proportional
with the ratio $Z_D/\sqrt{Q_{\alpha}}$.

The $\alpha$-decays between ground states are characterized by a remarkable
regularity, especially for transitions between ground states of even-even
nuclei.
The fact that $\alpha$-transitions along various isotopic chains lie on
separate lines, as stated by the Viola-Seaborg rule \cite{Vio66}, i.e.
\bea
\label{Viola}
log_{10}T=\frac{a_1Z_D+a_2}{\sqrt{Q_{\alpha}}}+b_1Z_D+b_2~,
\eea
is connected with different $\alpha$-particle reduced widths,
multiplying the penetrability in (\ref{gam}).
From Eq. (\ref{Tred}) it becomes clear that the reduced width should
depend upon the charge of the daughter nucleus.
This feature is shown in figure \ref{fig1}.
Still in doing systematics along neutron chains there are important
deviations with respect to this rule, as for instance in $\alpha$-decay
from odd mass nuclei, and this feature is strongly connected with
nuclear structure details.
Let us mention that different forms of the Viola-Seaborg relation were 
used in Refs. \cite{Hat90,Bro92}.

The Viola-Seaborg rule can be generalized for heavy-cluster decays 
\cite{Poe02}, as it is shown in figure \ref{fig2}. Here the angular 
momenta carried by emitted fragments are zero. 
In Ref. \cite{Ren04} it was proposed the following generalized
Viola-Seaborg rule for the heavy cluster emission
\bea
log_{10}T=a_1\frac{Z_DZ_C}{\sqrt{Q}}+a_2Z_DZ_C+b_2+c_2~,
\eea
with the following set of parameters
\begin{center}
$a_1=1.517~,~~~a_2=0.053~,~~~b_2=-92.911~,$
$c_2=0~(even),=1.402~(odd)~$,
\end{center}
where $c_2$ is the blocking parameter for odd-mass nuclei.
Thus, from Eq. (\ref{Tred}) the reduced width in this case
should depend upon the product between daugher and cluster charges.
 
\begin{figure}[ht] 
\begin{center} 
\includegraphics[width=9cm]{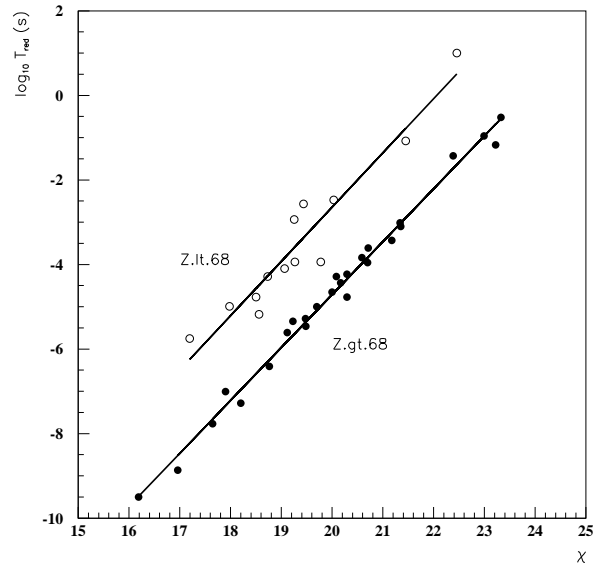} 
\vspace{-1cm} 
\caption{\it Logarithm of reduced half lives (\ref{Tred}) for proton 
emitters versus Coulomb parameter (\ref{Somer}).
Open circles denote emitters with $Z<68$, while dark circles
emitters with $Z>68$.}
\label{fig3} 
\end{center} 
\end{figure}

An interesting feature can be seen by plotting in figure \ref{fig3}
the logarithm of the reduced half life (\ref{Tred}) versus the Coulomb
parameter for various proton emitters \cite{Son02,Del06b}. 
In this case most of emitters have non-vanishing angular momentum.
The data are rougly divided into two regions, corresponding to
$Z<68$ (open circles) and $Z>68$ (dark circles).
This pattern can be asimilitated with a generalized Viola-Seaborg
rule for the two groups of charge numbers.

The situation with binary cold fission is quite different. 
Here there is not a simple linear dependence between the half-life for a 
given isotopic partition and the Coulomb parameter, due to the fact
that mass asymmetry changes during the scission process.

\section{A simple model for the reduced width}
\label{sec:model}
\setcounter{equation}{0} 
\renewcommand{\theequation}{3.\arabic{equation}} 

The decay processes can be schematically described
by the following cluster-daugher spherical potential
\bea
\label{simpo}
V(r)&=&\hbar\omega\frac{\beta (r-r_0)^2}{2}+v_0~,~~~r\leq r_B
\nn
    &=&\frac{Z_DZ_Ce^2}{r}\equiv V_C(r)~,~~~r>r_B~
\eea
where $r_0=1.2 A_D^{1/3}$ is the surface radius of the daugher nucleus.
Indeed, microscopic calculations have shown that the preformation
factor of the $\alpha$-particle has a Gaussian shape, centered
on the nuclear surface \cite{Del02}. Moreover, the spherical
component of the prefomation amplitude gives more than 90\% contribution
in the $\alpha$-decay width of deformed nuclei.
Notice that the radial equation of the shifted harmonic oscillator 
(ho) potential is similar with the equation of the one-dimensional
oscillator, but having approximate eigenvalues given by
\bea
E_{nl}=\hbar\omega\left(n+\oh\right)+\frac{\hbar^2l(l+1)}{2\mu r_0^2}~.
\eea
By considering $Q$-value as the first eigenstate in the shifted
ho well $Q-v_0=\oh\hbar\omega$,
together with the continuity condition at the top of
the barrier $r_B$, one obtains the following relation
\bea
\label{betar}
\hbar\omega\frac{\beta (r_B-r_0)^2}{2}=
V_{frag}(r_B)+\oh\hbar\omega~,
\eea
where it was introduced the so called fragmentation (or driving) 
potential, as the difference between the top of the Coulomb barrier and 
Q-value
\bea
\label{Vfrag}
V_{frag}(r_B)=V_C(r_B)-Q~.
\eea
The second component in Eq. (\ref{Tred}) contains the logarithm of the
Coulomb-Hankel function inside the Coulomb barrier which, according to 
Eq. (\ref{Hplus}), is proportional with the Colomb parameter $\chi$. 
The third part contains the reduced width squared which, according to 
Eq. (\ref{Pg}), is  proportional with the modulus of the internal wave 
function squared. 
For a shifted ho well one has for the ground state
\bea
|f^{(int)}_0(r)|^2=A_0^2e^{-\beta (r-r_0)^2}~.
\eea
By using the notation $\gamma\equiv\gamma_0$ and Eq. (\ref{betar}) 
one obtains the following relation
\bea
\label{gamfrag}
\log_{10}\gamma^2(r_B)=
-\frac{log_{10}e^2}{\hbar\omega}V_{frag}(r_B)
+\log_{10}\frac{\hbar^2A_0^2}{2e\mu r_B}~.
\eea

\begin{figure}[ht] 
\begin{center} 
\includegraphics[width=9cm]{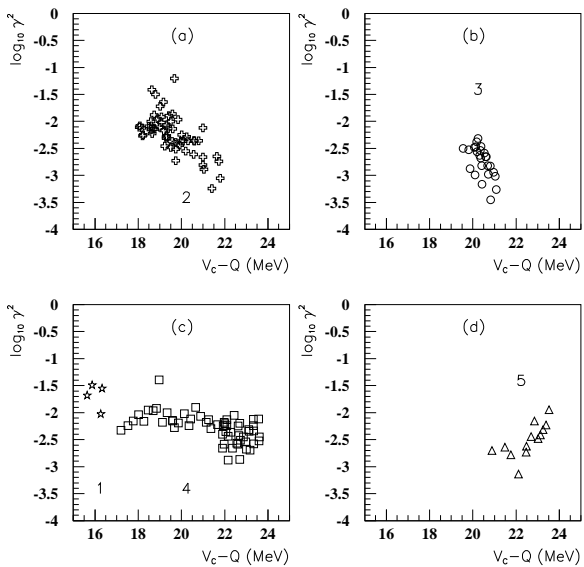} 
\vspace{-1cm}
\caption{\it 
The logarithm of the $\alpha$-decay reduced width squared versus the 
fragmentation potential (\ref{Vfrag}) for regions of the nuclear chart 
described by (\ref{zone}).} 
\label{fig4} 
\end{center} 
\end{figure}

In this way one obtains that indeed the logarithm of the half life
is of the Viola-Seaborg type
\bea
\label{Viola3}
\log_{10}T=c_1(r_B)\chi+c_2V_{frag}(r_B)+c_3(r_B,A_0^2)~,
\eea
because the fragmentation potential contains the product $Z_DZ_C$.
Its coefficient depends upon the touching radius, but this radius has a 
very small variation along various isotopic chains. 
Notice that the slope in Eq. (\ref{gamfrag}) has a negative value and
it is connected with the shape of the interaction potential (ho 
energy $\hbar\omega$), while the free term gives information about the 
amplitude of the cluster wave function.

Our calculation has shown that the linear relation (\ref{gamfrag})
but with different coefficients, remains valid
in the most general case of the double folding plus repulsive 
interaction between fragments, used in Refs. \cite{Pel07,Pel08}.

\section{Decay rule for reduced widths}
\label{sec:numer}
\setcounter{equation}{0} 
\renewcommand{\theequation}{4.\arabic{equation}} 

Most of experimental data refer to the $\alpha$-decay.
Therefore there were analyzed reduced widths in $\alpha$-decays
connecting ground states of even-even nuclei.
In figure \ref{fig4} it is plotted the logarithm of the experimental 
reduced width squared, by using the above relation versus the fragmentation
potential. The data are divided into five regions of even-even $\alpha$ 
emitters as follows

\bea
\label{zone}
1) & Z<82, ~~ 50<N<~82     & Fig.~\ref{fig4}~(c),~stars~;
\\
2) & Z<82, ~~ 82<N<126     & Fig.~\ref{fig4}~(a),~crosses~;
\nn
3) & Z>82, ~~ 82<N<126     & Fig.~\ref{fig4}~(b),~circles~; 
\nn
4) & Z>82, ~ 126<N<152     & Fig.~\ref{fig4}~(c),~squares~;
\nn
5) & Z>82, ~~~~~~~~~ N>152 & Fig.~\ref{fig4}~(d),~triangles~.
\nonumber
\eea

In calculations it was used the value of the touching radius, i.e.
\bea
\label{radius}
r_B=1.2(A_D^{1/3}+A_C^{1/3})~,
\eea
where the approximation $K_{ll'}\approx\delta_{ll'}$ in Eq. (\ref{Kll}) 
is fulfilled with $90\%$ accuracy for the most deformed nuclei 
\cite{Del06}.
Notice that regions 1-4 contain rather long isotopic chains,
while in the last region 5 one has not more than two isotopes/chain.
This is the reason why, except for the last region 5, the reduced width 
decreases with respect to the fragmentation potential,
according to the theoretical prediction given by (\ref{gamfrag}).
Notice that in the regions 1 and 4, above $^{100}$Sn and
$^{208}$Pb double magic emitters, respectivelly, the ho parameter of the 
$\alpha$-daughter potential is larger than for regions 2 and 3,
corresponding to charge numbers around the double magic nucleus 
$^{208}$Pb.
On the other hand, one obtains the largest amplitudes
of the cluster wave function in the regions 2 and 3.

\begin{figure}[ht] 
\begin{center} 
\includegraphics[width=9cm]{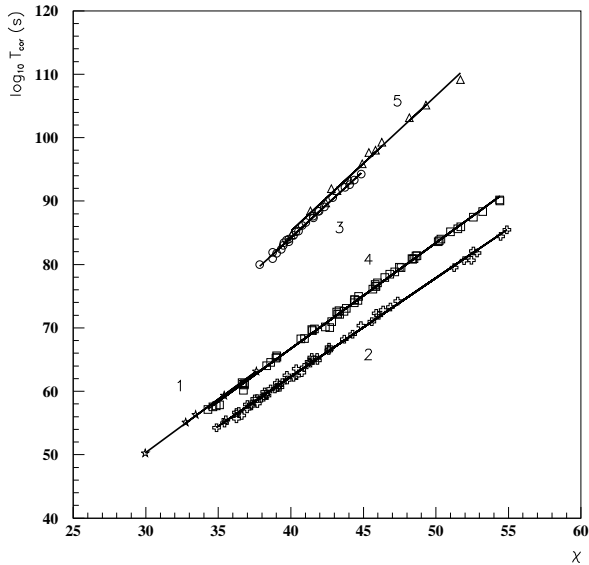} 
\vspace{-1cm}
\caption{\it 
The difference $\log_{10}T-c_2V_{frag}(r_B)-c_3$ 
versus the Coulomb parameter $\chi$ for five different regions described
by (\ref{zone}).
The straight lines are the corresponding linear fits.
} 
\label{fig5} 
\end{center} 
\end{figure}

An interesting observation can be made from figure \ref{fig5}, where it 
is plotted the difference $\log_{10}T-c_2V_{frag}(r_B)-c_3(r_B,A_0^2)$
versus the Coulomb parameter $\chi$,
by using the same five symbols for the above described regions.
Amazingly enough there were obtained three lines, corresponding to 
different amplitudes of the cluster wave function.
The regions 1 and 4, corresponding to emitters above double magic nuclei
$^{50}$Sn and $^{208}$Pb, respectivelly, have practically the same internal
amplitudes $A_0$. The same is true for the regions 3 and 5.

\begin{figure}[ht] 
\begin{center} 
\includegraphics[width=9cm]{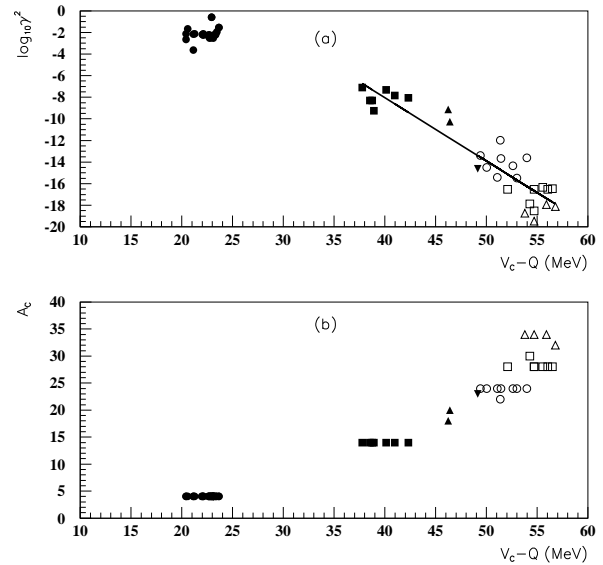} 
\vspace{-1cm}
\caption{\it 
(a) The logarithm of the reduced width squared versus the fragmentation 
potential (\ref{Vfrag}). Different symbols correspond to cluster decays 
in figure \ref{fig2}. The straight line is the linear fit 
(\ref{fitclus}) for cluster emission processes, except $\alpha$-decay.
(b) Cluster mass number versus the fragmentation potential.}
\label{fig6} 
\end{center} 
\end{figure}

The linear dependence of $log_{10}\gamma^2$ versus the fragmentation 
potential (\ref{gamfrag}) remains valid for any kind
of cluster emission. This fact is nicely confirmed by heavy
cluster emission processes in figure \ref{fig6} (a), where it is 
plotted the dependence between the corresponding experimental values
for the same decays in figure \ref{fig2}. Here it is also plotted
a similar dependence for $\alpha$-decays corresponding to
the same heavy cluster emitters.
The straight line is the linear fit for cluster emission processes,
except $\alpha$-decays 
\bea
\label{fitclus}
log_{10}\gamma^2=-0.586 (V_C-Q)+15.399~.
\eea

The above value of the slope $-log_{10}e^2/\hbar\omega$ in Eq. 
(\ref{gamfrag}) leads to $\hbar\omega\approx 1.5$ MeV, with
the same order of magnitude as in the $\alpha$-decay case.
The relative large scattering of experimental data around the
straight line in figure \ref{fig6} can be explained by the simplicity
of the used cluster-core potential (\ref{simpo}).

Let us mention that a relation expressing the spectroscopic factor
(proportional with the reduced width) for cluster emission processes
was derived in Ref. \cite{Ble91}
\bea
S=S_{\alpha}^{(A_C-1)/3}~,
\eea
where $A_C$ is the mass of the emitted light cluster and
$S_{\alpha}\sim 10^{-2}$. As can be seen from figure \ref{fig6} (b)\ 
between $A_C$ and $V_{frag}$ there exists a rather good linear 
dependence and therefore the above scalling law can be easily understood 
in terms of the fragmentation potential.

\begin{figure}[ht] 
\begin{center} 
\includegraphics[width=9cm]{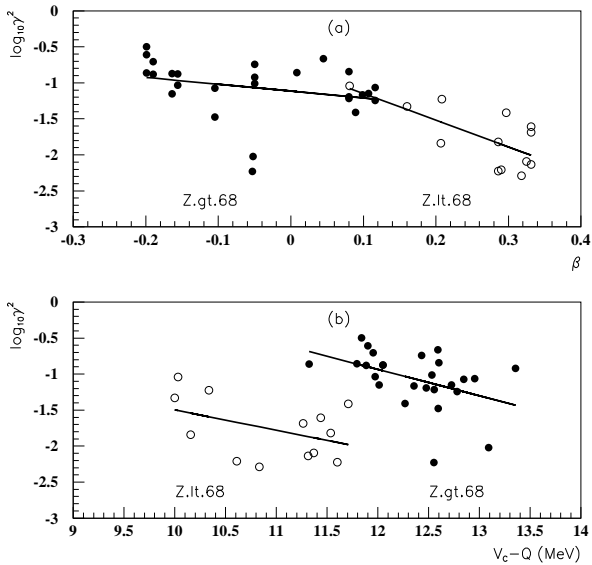} 
\vspace{-1cm}
\caption{\it 
(a) The logarithm of the reduced width squared versus the
quadrupole deformation.
By open circles are given emitters with $Z<68$, while by dark circles 
those with $Z>68$ for proton emission.
The two regression lines fit the corresponding data.
(b) The logarithm of the reduced width squared versus the 
fragmentation potential (\ref{Vfrag}). The symbols are
the same as in (a).}
\label{fig7} 
\end{center} 
\end{figure}

Concerning the reduced widths of proton emitters in Refs. 
\cite{Del06b,Med07} it was pointed out the correlation
between the reduced width and the quadrupole deformation.
This fact can be seen in figure \ref{fig7} (a), where the
region with $Z<68$ corresponds to $\beta>0.1$ (open circles), while
the other one with $Z>68$ to $\beta<0.1$ (dark circles). The two
linear fits have obviously different slopes.
This dependence is induced by the propagator matrix $K_{ll'}(\beta,r)$
in Eq. (\ref{Kll}).
Notice that the two dark circles with the smallest reduced 
widths correspond to the heaviest emitters with $Z>80$.

At the same time one sees from figure \ref{fig7} (b)
that the same data are clustered into two regions, which can
be directly related with the fragmentation potential (\ref{Vfrag}).
Here the two linear fits in terms of the fragmentation 
potential, corresponding to the two regions
of charge numbers, have roughly the same slopes, but different
values in origin. Thus, the two diferrent lines in figure \ref{fig3}
can be directly connected with similar lines in figure \ref{fig7} (b). 
They correspond to different orders of magnitude of the
fragmentation potential, giving different orders to wave functions
and therefore to reduced widths.

\begin{figure}[ht] 
\begin{center} 
\includegraphics[width=9cm]{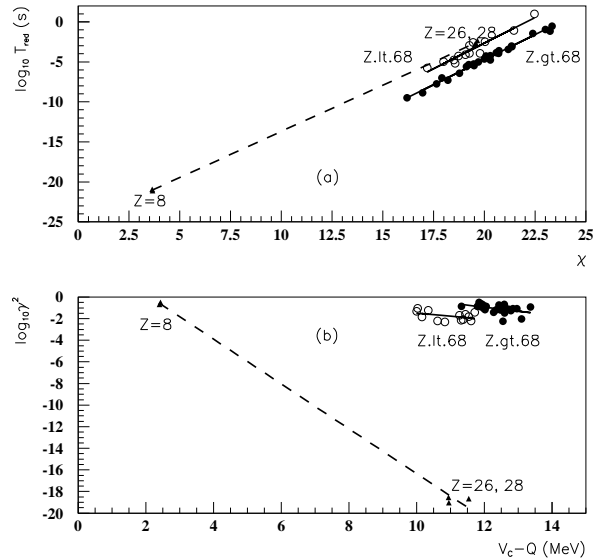} 
\vspace{-1cm}
\caption{\it 
(a) The logarithm of the half life versus Coulomb parameter
for two-proton emitters (triangles). By circles are
given data for proton emitters in figure \ref{fig3}.
(b) The logarithm of the reduced width squared versus the 
fragmentation potential (\ref{Vfrag}) for two-proton emitters 
(triangles).
The same quantity is given by circles for proton emitters
in figure \ref{fig7} (b).
The solid lines fit proton data, while the dashed lines
fit two-proton data.}
\label{fig8} 
\end{center} 
\end{figure}

A special case is given by the two-proton emission. 
This process was predicted log time ago \cite{Gol60},
but only few such emitters were recently detected until now.
Let us mention that the most recent general treatment of the two-proton 
emission process, assuming a three-body dynamics, is  given in Refs. 
\cite{Gri07} (and the references therein).

The experimental half-lives versus Coulomb parameter are given in 
figure \ref{fig8} (a) by triangles. 
The emitter charges are also pointed out.
Here it is assumed a simplified version, where the light emitted 
cluster is supposed to be a di-proton with the charge $Z_C=2$. In this 
case one can use the factorisation of the decay width (\ref{gam}). 
One sees that half lives (triangles) follow the 
general trend (dashed line) of the usual proton emitters, given in the 
same figure by the symbols in figure \ref{fig3}. 
In figure \ref{fig8} (b) it is plotted the logarithm of the reduced
width squared versus the fragmentation potential by triangles.
One indeed observes that the slope of the fitting dashed line
has a negative value $-log_{10}e^2/\hbar\omega$, but with a
much larger value in comparison with proton emitters, given by the same 
symbols as in figure \ref{fig7} (b). 
The three lines in this figure, corresponding to
proton and two-proton radioactivity, respectivelly, are given by
\bea
log_{10}\gamma^2&=&-0.283(V_C-Q)+1.329~,~Z<68
\nn
log_{10}\gamma^2&=&-0.365(V_C-Q)+3.440~,~Z>68
\nn
log_{10}\gamma^2&=&-2.075(V_C-Q)+4.403~.
\eea
The ho energy is $\hbar\omega\approx 1.5$ MeV for proton emission 
(i.e. the same order as for heavy cluster radioactivity and
$\alpha$-decay) and
$\hbar\omega\approx 0.2$ MeV for two-proton emission,
by considering the di-proton approximation.
This small ho energy is connected with the use of Eq. (\ref{radius})
in estimating the spatial extension of the di-proton system $R_C$
and therefore the fragmentation potential.
In reality the di-proton is not a bound system and it changes its
size during the barrier penetration. 
Our microscopic estimate, by using a pairing residual interaction 
between emitted protons from $^{45}$Fe, evidenced that the 
size of the wave packet increases in the relative coordinate by 1 fm 
over a distance of 1 fm in the region of the nuclear surface.
Actually the reduced width defined by (\ref{gam}) can be generalized by 
using the other extreme scenario, given by a sequential emission, where 
this relation is integrated over all possible energies of emitted 
particles \cite{Kry95}.

An interesting observation concerns the amplitude $A_0$ in Eq. 
(\ref{gamfrag}), given by the value of fitting lines in origin. 
It has similar values for both two-proton and proton emitters
with $Z>68$.

Now let us analyze $\alpha$-decay processes to excited 
low lying $2^+$ states.
There were considered more than 70 decays of even-even rotational, 
vibrational and transitional nuclei \cite{Pel07,Pel08}.
The hindrance factor (HF) is defined as the ratio of reduced widths 
squared connecting the ground states and ground to excited states with
the angular momentum $l=2$, i.e.
\bea
HF=\frac{\gamma^2_0}{\gamma^2_2}~.
\eea
Thus, by using (\ref{gamfrag}), the logarithm of the HF
becomes proportional with the excitation energy of the daughter nucleus
\bea
\label{lgHF}
log_{10}HF=\frac{log_{10}e^2}{\hbar\omega}E^{(ex)}_2
+log_{10}\frac{A_0^2}{A_{2}^2}~.
\eea
It is worth mentioning that this relation is equivalent
with the Boltzman distribution for the reduced width 
to the excited state $\gamma_2$. 
In Refs. \cite{Xu07,Wan09} such a dependence was postulated
in order to  describe HF's.

\begin{figure}[ht] 
\begin{center} 
\includegraphics[width=9cm]{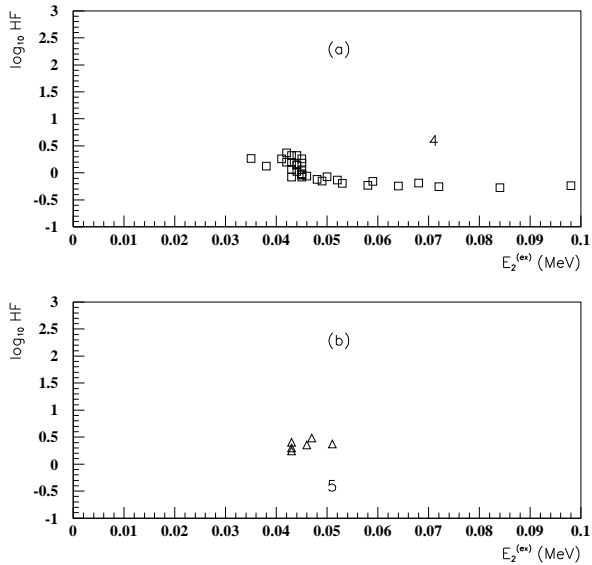} 
\vspace{-1cm}
\caption{\it 
The logarithm of the hindrance factor versus the excitation
energy of the daughter nucleus for rotational nuclei.
The symbols and numbers correspond to the regions given by Eq. 
(\ref{zone}).}
\label{fig9} 
\end{center} 
\end{figure}

\begin{figure}[ht] 
\begin{center} 
\includegraphics[width=9cm]{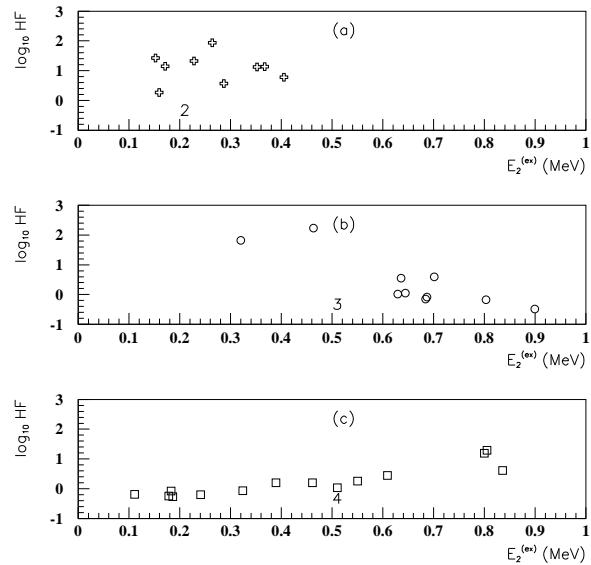} 
\vspace{-1cm}
\caption{\it 
The same as in figure \ref{fig9}, but for transitional
and vibrational nuclei.}
\label{fig10} 
\end{center} 
\end{figure}

In figure \ref{fig9} it is plotted the logarithm of the HF
versus the excitation energy for rotational nuclei with 
$E_{2}^{(ex)}<$ 0.1 MeV, by using the same notations
given by Eq. (\ref{zone}).
As a rule the HF's have small values and therefore the wave functions 
have  similar amplitudes $A_0\approx A_2$. 
Notice that the region 4 in figure \ref{fig9} (b), with 
$Z>82,~126<N<126$, contains most of rotational emitters (33 out of 41). 
Moreover, our analysis has shown that here the HF has an almost constant
value along various isotopic chains, due to the fact that the energy 
range is very short (about 100 keV).

In figure \ref{fig10} it is given the same quantity, but for 
transitional and vibrational nuclei, with $E_{2}^{(ex)}>$ 0.1 MeV. 
The situation looks here to be different and more complex, with
respect to rotational nuclei.
The best example is given by the same region 4 in figure 
\ref{fig10} (c), where the slope has a positive value, as predicted by
Eq. (\ref{lgHF}).
Notice that this region contains almost half of the analyzed 
vibrational emitters (14 out of 31).

\section{Conclusions}
\label{sec:concl}
\setcounter{equation}{0} 
\renewcommand{\theequation}{5.\arabic{equation}} 

Cold emission processes in terms of the well known factorisation of the 
decay width between the penetrability and reduced width squared, were 
analyzed.
Based on a simple model of the two-body dynamics,
namely a shifted harmonic oscillator potential surounded by the Coulomb 
interaction, it was derived an universal analytical relation expressing 
the logarithm of the reduced width squared as a linear function in terms of 
the  fragmentation potential, defined as the difference between the 
Coulomb barrier and the Q-value. Notice that the slope  has the same 
order of magnitude, corresponding to an ho energy 
$\hbar\omega\approx 1.5$ MeV,  for all decay processes, except the 
di-proton emission, where the factorisation (\ref{gam}) is
not anymore valid.
This rule is a consequence of the fact that the logarithm 
of the wave function squared is proportional with the difference
between the height of the Coulomb potential at a given radius and the 
energy of the system.
It is fulfilled with a reasonable accuracy by experimental data, 
describing transitions between  ground states as well as for 
$\alpha$-transitions to  excited states in the most relevant region 
with $Z>82,~126<N<152$.
As a particular case, the two regions in the dependence of proton 
emitters half-lives, corrected by the centrifugal barrier, versus
the Coulomb parameter are directly related with the corresponding 
regions of the fragmentation potential.
Thus, the clustering character of the wave function inside the Coulomb
barrier (i.e. the fragments are already preformed here) is evidenced by 
this linear rule in terms of the fragmentation potential.
On the other hand the well known scalling relation, connecting the
spectroscopic factor with the mass of the emitted cluster, can
be nicely explained in terms of the fragmentation potential.

\section{Acknowlegments}
\label{sec:acknow}
\setcounter{equation}{0} 
\renewcommand{\theequation}{6.\arabic{equation}} 

This work was supported by the contract IDEI-119 of the Romanian
Ministry of Education and Research.


\end{document}